\documentclass[aps,prd,amsmath,amssymb,nofootinbib,twocolumn,showpacs]{revtex4}
\usepackage{bm}
\usepackage{bbm}
\usepackage{color}
\usepackage{graphicx}
\usepackage{amsmath,amssymb}

\newcommand{\cont}{\text{cont}}
\newcommand{\Nt}{N_\text{t}}
\newcommand{\Ns}{N_\text{s}}
\newcommand{\SB}{S_\text{B}}
\newcommand{\mlatt}{m_\text{latt}}

\newcommand{\tskip}{t_\text{skip}}
\newcommand{\meff}{m_\text{eff}}

\DeclareMathOperator{\diag}{diag}

\newcommand{\half}{{\textstyle\frac{1}{2}}}

\newcommand{\abs}[1]{\left| #1 \right|}
\newcommand{\vev}[1]{\left\langle #1 \right\rangle}
\newcommand{\ord}{\mathcal{O}}

\graphicspath{{./figures/}}
\begin{document}
\title{Two-Dimensional Wess-Zumino Models at Intermediate Couplings}

\author{Tobias K\"astner}
\author{Georg Bergner}
\author{Sebastian Uhlmann}
\author{Andreas Wipf}
\author{Christian Wozar}
\thanks{\texttt{T.Kaestner}, \texttt{G.Bergner}, \texttt{A.Wipf@tpi.uni-jena.de}
and \texttt{S.Uhlmann}, \texttt{Christian.Wozar@uni-jena.de}}
\affiliation{Theoretisch-Physikalisches Institut,
Friedrich-Schiller-Universit{\"a}t Jena, Max-Wien-Platz 1, 07743
Jena, Germany}


\begin{abstract}
\noindent
We consider the two-dimensional ${\cal N}=(2,2)$ Wess-Zumino model with a cubic
superpotential at weak and intermediate couplings. Refined algorithms allow
for the extraction of reliable masses in a region where perturbation theory
no longer applies. We scrutinize the Nicolai improvement program which is
supposed to guarantee lattice supersymmetry and compare the results for ordinary
and non-standard Wilson fermions with those for SLAC derivatives. It turns out
that this improvement completely fails to enhance simulations for Wilson fermions
and only leads to better results for SLAC fermions. Furthermore, even without
improvement terms the models with all three fermion species reproduce the correct
values for the fermion masses in the continuum limit.
\end{abstract}

\maketitle

\section{Introduction}
\noindent
Supersymmetric models have drawn much attention over the past decades.
In particular, supersymmetric extensions of the standard model have become
a primary research topic for model building. The additional symmetry of
these models proves to be a very useful tool for the study of their
perturbative and non-perturbative aspects. It is notoriously complicated
to check and extend the predictions made by supersymmetry in a strong
coupling regime where standard perturbation no longer applies.

At the same time, lattice simulations of quantum field theories have been
very successful in an increasing number of applications. In some theories,
it is possible to match numeric results at weak coupling to perturbative
continuum results; at stronger coupling, lattice simulations are often the
only viable way to investigate non-perturbative properties of the theories.
As nonperturbative effects are automatically taken into account, it is
desirable to apply the lattice approach also to supersymmetric theories.
This has been the subject of a number of publications, see, e.g.,
\cite{Feo:2004kx, Giedt:2006pd} and references therein.
There are a number of challenges with respect to this goal since it is well
known that full supersymmetry can not be realized in a generic lattice model.
The reason for this can be traced back to the failure of the
Leibniz rule on the lattice~\cite{Dondi:1976tx}. The full supersymmetry can
only be recovered in the limit of vanishing lattice spacing (continuum limit);
but only in some cases, the conditions for such a restoration are under control. 
E.g., it has been shown that even in supersymmetric quantum mechanics the naive
discretization does not lead to a supersymmetric continuum limit~\cite{Giedt:2004qs};
generically, such a limit can at best be achieved by finetuning the bare
coefficients of all supersymmetry-breaking counterterms~\cite{Montvay:1995rs}.
This, however, requires much knowledge of the theory in advance.
In some cases the relevant operators can be determined perturbatively,
cf.~\cite{Golterman:1988ta}. A possible way beyond perturbation theory is the
application of a blocking transfromation as in~\cite{Bietenholz:1998qq} for
a free theory. This may lead to a solution similar to the
Ginsparg-Wilson relation for the chiral symmetry~\cite{Bergner:2008ws}.

It is possible to reduce the number of relevant operators in the continuum 
limit if some symmetries of the continuum theory are already realized in the 
lattice action. The standard lore is that it is sufficient to
realize just a part of the supersymmetry on the lattice in order to ensure the 
correct continuum limit. There have been many suggestions and numerical 
investigations with respect to such a partial realization of the supersymmetry 
algebra on the lattice, e.g.\ \cite{Elitzur:1982vh} and \cite{Kaplan:2002wv}. 
An elegant suggestion uses a Nicolai map~\cite{Nicolai:1979nr} to create
lattice improvement terms that guarantee a partial realization of supersymmetry,
cf.\ e.g.\ \cite{Catterall:2001fr}.

Numerical simulations of supersymmetric theories face the further difficulty
that bosons and fermions on the lattice should be treated on equal footing.
This demands for dynamical fermions; however, such simulations are notoriously
numerically involved. Therefore, it is advisable to start with low-dimensional
theories in order to gain information about the performance of the different
supersymmetric lattice formulations. On the other hand, such dynamical fermion
simulations in low dimensions are interesting in their own right because they
allow for an explicit investigation and improvement of the corresponding known
algorithms.

We have started the analysis of such low-dimensional models in a previous paper~%
\cite{Bergner:2007pu} with investigations of various lattice formulations of
supersymmetric quantum mechanics and first tests of the two-dimensional
Wess-Zumino model at weak coupling. Here we will extend the analysis of the
latter theory using far more elaborate numerical techniques to reach 
intermediate to strong values of the coupling. We are able
to simulate the Wess-Zumino model for a much larger parameter
region as in related previous works~\cite{Catterall:2001fr}
and~\cite{Beccaria:1998vi}. Starting from the standard
hybrid Monte Carlo algorithm~\cite{Duane:1987de} we employ a novel
combination of algorithms involving both a higher-order~\cite{omelyan:2003} integration scheme and
Fourier acceleration~\cite{Weingarten:1980hx}. This entails
much better statistics in combination with larger lattice sizes. These
improvements lead to reliable new results even at stronger coupling where
considerable deviations from perturbative predictions, e.g., for the masses of the supersymmetric
partners can be observed.

A further goal was a systematic study of the effects of the above-mentioned
improvement terms introduced by the Nicolai map~\cite{Catterall:2001fr}. In
this paper, we present the first explicit comparison of the models with and
without such terms. It may come as a surprise that for Wilson fermions the
``improvement term'' even fails to improve the properties of the lattice
model. Moreover, such terms introduce new complications and can lead to
unreliable numerical results.

In previous works~\cite{Bergner:2007pu, Kirchberg:2004vm} it has been 
demonstrated that lattice models based on the SLAC derivative~\cite{Drell:1976mj} and on the
twisted Wilson formulation (as introduced in \cite{Bergner:2007pu}) are
particularly well-behaved as far as the continuum limit is concerned. Even
at large lattice spacing the continuum result is approximated very well. In
the current simulation the SLAC derivative again proves to be the best choice
because it allows for much larger values of the coupling constant, and only a
comparably coarse lattice is needed to extract the correct continuum results.
It is interesting to note that contrary to a realization with Wilson fermions
the improvement terms for the SLAC derivative in fact lead to better numerical
results.

The paper is organized as follows: We start out with a short introduction of
the different lattice realizations of the two-dimensional ${\cal N}=2$ Wess-Zumino
model and the corresponding improvement terms with their respective
lattice and continuum symmetries. Then, we present the numerical results
of our simulations; in particular, we compare the masses of the supersymmetric
partners as a measure for how well supersymmetry is realized on the lattice.
A comparison of the various models with the perturbative continuum prediction
at smaller values of the dimensionless coupling is the subject of Section~\ref{ssec:weakCoupling}.
At last, we turn special attention to the regime of intermediate couplings where the
measured masses differ considerably from the one-loop results.

\section{Lattice models}

\subsection{Supersymmetrically improved lattice actions}
\label{sec:wz2d:complex}
\noindent
The lattice models under consideration have been
discussed at length in \cite{Bergner:2007pu}. Therefore, we
shall only briefly recall the definitions of the corresponding lattice
actions. In terms of complex coordinates $z$ and $\bar z$ for the
two-dimensional Euclidean spacetime together
with the corresponding holomorphic and anti-holomorphic differentials
$\partial$ and $\bar\partial$ the continuum action of the ${\cal N} = 2$
Wess-Zumino model reads
\begin{equation}
\label{eq:wz2d:nic1}
S_\text{cont} = \int d^2x
\left(
2\bar\partial\bar\varphi\partial\varphi+
\half |W'(\varphi)|^2 +\bar\psi M\psi
\right).
\end{equation}
The bosonic potential is given by the absolute square of the derivative
of the holomorphic superpotential $W(\varphi)$ w.r.t.\ its argument
$\varphi=\varphi_1+i\varphi_2$.
Apart from the standard kinetic term for the (two-component) Dirac
spinors, the Dirac operator $M$ contains a Yukawa coupling,
\begin{equation}
\label{eq:wz2d:nic2}
M=\gamma^z\partial + \gamma^{\bar z}\bar\partial + W''P_{+}
+\overline{W}''P_{-}.
\end{equation} 
In \eqref{eq:wz2d:nic2} we have introduced chiral
projectors $P_{\pm}=\half(\mathbbm{1}\pm\gamma_3)$
which in the Weyl basis with $\gamma^1=\sigma_1$,
$\gamma^2=-\sigma_2$, $\gamma_3=i\gamma^1\gamma^2$ project
onto the upper and lower components of $\psi$.
In the form \eqref{eq:wz2d:nic1} the action is invariant
under four real supercharges. Taken together they satisfy the 
$\mathcal N=(2,2)$ superalgebra, and it has been argued
that at most one supersymmetry can be preserved on the lattice \cite{Catterall:2001fr}.
With the help of the explicitly known form of the Nicolai map it is
possible to construct such a lattice model straightforwardly.
In terms of the Nicolai variable
$\xi_x=2(\bar\partial\bar\varphi)_x+W_x$ on the lattice, the
discretized Wess-Zumino action reads
\renewcommand{\arraystretch}{1.4}
\begin{equation}
\label{eq:wz2d:nic6}
S = \half\sum_x \bar\xi_x\xi_x + \sum_{x,y}\bar\psi_x M_{xy}
\psi_y.
\end{equation}
Here, $W_x$ is taken to be the lattice counterpart of the 
continuum operator $W'(\varphi)$, i.e.\ $W_x=W'(\varphi_x)$. 
The matrix $M$ is given by
\begin{equation} 
\label{eq:wz2d:nic7}
M_{xy} = \begin{pmatrix}
         W_{xy} & 2\bar\partial_{xy}\\
         2\partial_{xy} & \overline{W}_{xy} 
         \end{pmatrix}
= \begin{pmatrix}
         \frac{\partial \xi_x}{\partial\varphi_y}& 
         \frac{\partial \xi_x}{\partial\bar\varphi_y} \\
         \frac{\partial \bar\xi_x}{\partial\varphi_y}&  \frac{\partial \bar\xi_x}{\partial\bar\varphi_y}  
         \end{pmatrix}.
\end{equation}
\renewcommand{\arraystretch}{1.0}%
We require all lattice difference operators to be antisymmetric,
$\partial_{xy}=-\partial_{yx}$. 
From the second equality in \eqref{eq:wz2d:nic7} we can read off
that $W_{xy}:=\partial W_x / \partial\varphi_y$.

One easily checks that \eqref{eq:wz2d:nic6} is
invariant under the following (supersymmetry) variation,
\begin{subequations}
\label{eq:wz2d:nic8}
\begin{align} 
\delta\varphi_x &=\bar\varepsilon\psi_{1,x}  ,& 
\delta\bar\psi_{1,x} &= -\half\bar\xi_x\bar\varepsilon,& 
\delta\psi_{1,x} &= 0, \\
\delta\bar\varphi_x &=\bar\varepsilon\psi_{2,x}, & 
\delta\bar\psi_{2,x} &= -\half\xi_x\bar\varepsilon,& 
\delta\psi_{2,x} &= 0.
\end{align}
\end{subequations}
In terms of the original fields, \eqref{eq:wz2d:nic6} takes the form
\begin{equation} 
\label{eq:wz2d:nic11}
\begin{split}
S\!=\!&\sum_x \!\Big(\!2\!\left(\bar\partial\bar\varphi\right)_x(\partial\varphi)_x + \half\big|W_x\big|^2 
+ W_x(\partial\varphi)_x + \overline{W}_{\!x}(\bar\partial\bar\varphi)_x\!\Big)\\ 
&+\sum_{x,y}(\bar\psi_{1,x}, \bar\psi_{2,x}) 
\begin{pmatrix}
W_{xy} & 2\bar\partial_{xy}\\
2\partial_{xy} & \overline{W}_{xy}
\end{pmatrix}
 \begin{pmatrix}
 \psi_{1,y}\\\psi_{2,y}
 \end{pmatrix}.
\raisetag{1.7em}
\end{split}
\end{equation}
This supersymmetrically improved lattice action differs from a
straightforward discretization of \eqref{eq:wz2d:nic1} by
\begin{equation}
\label{eq:wz2d:nic12}
\Delta S = \sum_x \Big(W_x(\partial\varphi)_x+
\overline{W}_x(\bar\partial\bar\varphi)_x\Big)
\end{equation}
a discretization of a surface term in the continuum theory (which is
therefore expected to vanish in the continuum limit for suitably chosen
boundary conditions). For the free theory
($W_x=m\varphi_x$) $\Delta S =0$ readily follows from the
antisymmetry of the difference operator $\partial_{xy}$
while for interacting theories \eqref{eq:wz2d:nic12}
guarantees the invariance of the action under
\eqref{eq:wz2d:nic8} without the need of the Leibniz rule. 
To study the impact of SUSY improvement we will compare
also the improved action with the \emph{unimproved} straightforward
discretization of \eqref{eq:wz2d:nic1} (without $\Delta S$).

\subsection{Lattice fermions}
\noindent
For the symmetric difference operator
\begin{equation}
\label{eq:nic13}
\left(\partial^S_{\mu}\right)_{xy}=\half(\delta_{x+\hat\mu,y}-\delta_{x-\hat\mu,y}),
\end{equation}
doublers are inevitably introduced into both the bosonic
and fermionic sector. In order to get rid of them a Wilson
term may be added to the superpotential so as to maintain the
invariance of the action under \eqref{eq:wz2d:nic8}.
Within this context two different choices have been
discussed previously~\cite{Bergner:2007pu},
\begin{equation}
\label{eq:stdWilson}
W_x = W'(\varphi_x) - \frac{r}{2} (\Delta\varphi)_x
\end{equation}
and
\begin{equation}
\label{eq:twistWilson}
W_x = W'(\varphi_x) + \frac{i r}{2} (\Delta\varphi)_x.
\end{equation}
We stress that for Wilson fermions, the derivative of the superpotential
is now shifted as compared to the situation after~\eqref{eq:wz2d:nic6}.
From the first expression we recover the standard Wilson
term for the fermions, i.e.\
$W_{xy}=W''(\varphi_x)\delta_{xy} -
\frac{r}{2}\Delta_{xy}$.
The operator $\Delta_{xy}$ is the usual two-dimensional
(lattice) Laplacian $2\partial\bar\partial$.
The second possibility \eqref{eq:twistWilson} leads to $W_{xy}=W''(\varphi_x)\delta_{xy} +
\gamma_3\frac{r}{2}\Delta_{xy}$. Here, the appearance of
$\gamma_3$ motivates the name \emph{twisted} Wilson fermions
(not be confused with the recently introduced twisted mass formulation of lattice
QCD). It was already shown for the free theory~\cite{Bergner:2007pu} that
\emph{twisted} Wilson fermions suffer far less from lattice artifacts
than their standard Wilson cousins. Here we will show
that they remain superior even for (strongly) interacting theories.

Besides these two (ultra-)local difference operators we
have previously suggested to reconsider the non-local
SLAC lattice derivative in the context of lattice
Wess-Zumino models. The matrix elements of the SLAC derivative are
most conveniently given for a one-dimensional lattice with
an odd number of lattice points $L$,
\begin{equation}
\label{eq:qm:slac}
\partial_{x\neq
y}=(-1)^{x-y}\frac{\pi/L}{\sin(\pi(x-y)/L)}, \quad
\partial_{xx}=0.
\end{equation}
The generalization to higher dimensions is
straightforward and amounts to forming suitable tensor
products of \eqref{eq:qm:slac}.%
\footnote{The reason for an odd
number of lattice points originates from a reality
condition on the matrix elements \eqref{eq:qm:slac}. As such it is
a mere technicality in order to ease numerical simulations.}
For SLAC fermions no further modifications to the
superpotential are necessary. It is due to this fact that
they constitute an interesting alternative to Wilson
fermions.

\subsection{Discrete symmetries} \label{sec:wz2d:class}
\noindent
For the numerical analysis of Sec.~\ref{sec:numresults} we
have chosen the superpotential 
\begin{equation}
\label{eq:superpot}
W(\varphi) =
\half m\varphi^2+\textstyle{\frac{1}{3}}g\varphi^3
\end{equation}
which coincides with that in earlier simulations
of the Wess-Zumino model~\cite{Beccaria:1998vi, Catterall:2001fr}.
\begin{figure}
  \input{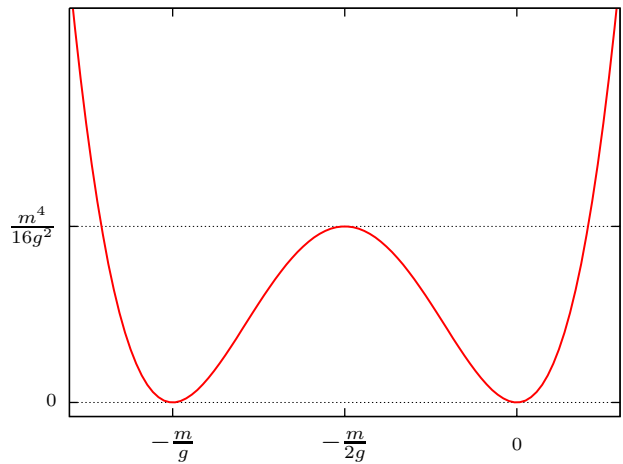}%
  \caption[labelInTOC]{Classical bosonic potential
  $V(\varphi) = \half|W'(\varphi)|^2$ from
  \eqref{eq:superpot} shown for vanishing imaginary part
  ($\varphi_2=0$). In the free theory limit ($g\to 0$) the
  left minimum is pushed towards minus infinity.}
  \label{fig:classPot}
\end{figure}
We will assume the coupling constants $m$ and $g$ to be
real and positive.
The superpotential \eqref{eq:superpot} allows for discrete symmetries
$\mathbb Z_2^\text{R}\times \mathbb Z_2^\text{C}$ which act as reflections interchanging
the two vacua and as complex conjugations on the complex scalar field:
\begin{equation} 
\label{eq:wz2d:symm1}
\mathbb Z_2^\text{R}\colon\varphi \mapsto - \frac{m}{g} -\varphi\quad
\text{and}\quad\mathbb Z_2^\text{C}\colon\varphi\to \bar\varphi,
\end{equation}
so that also the potential $\frac{1}{2}|W'(\varphi)|^2$ is invariant
under both transformations, cf.~Fig.~\ref{fig:classPot}.

From the explicit form of the fermion matrix $M$ and its
adjoint~$M^\dagger$ 
\begin{subequations}
\begin{align}
\label{eq:wz2d:symm4}
M&=\gamma^\mu\partial_\mu  + m + 2g(\varphi_1 + i\gamma_3\varphi_2),\\
M^\dagger&=-\gamma^\mu\partial_\mu  + m + 2g(\varphi_1 - i\gamma_3\varphi_2)
\end{align}
\end{subequations}
one finds that
\begin{equation}
\label{eq:wz2d:symm6}
\mathbb Z_2^\text{R}\colon M \mapsto -\,\gamma_3M\gamma_3,\quad
\mathbb Z_2^\text{C}\colon M \mapsto  \gamma_3M^\dagger\gamma_3,
\end{equation}
which shows the invariance of the determinant.%
\footnote{This is true at least up to an irrelevant sign.
On the lattice the fermion matrix
$M$ always has an even number of rows and columns, hence
this phase does not appear.} %
\begin{table*}
\caption{\label{tab:wz2d:fermions}Comparison of 
various lattice models w.r.t.\ their symmetries. All statements refer to to the interacting theory,
i.e.\ $g\neq 0$. The notion $\mathbb Z_2^\text{PC}$
denotes the combined action of a field and parity transformation as discussed in the text.}
\begin{ruledtabular}
\begin{tabular}{lccccc}
&(1)&(2)&(3)&(4)&(5)\\
					& Wilson impr. & Wilson unimpr. & twisted Wilson\footnote{Only improved considered.} & SLAC impr. &  SLAC unimpr.\\\hline
lattice derivative  & local	 & local & local          & non-local & non-local \\
lattice  artifacts	& $\ord(a)$ & $\ord(a)$ & $\ord(a)$%
\footnote{In the interacting case the good scaling properties are lost. However the overall size of lattice artifacts is still much smaller when
compared to Wilson fermions.}%
 & `perfect' & `perfect'\footnote{The dispersion relation is up to 
the cut-off the same as in the continuum.}\ \\ modifications to superpot.  & yes
    & yes	  & yes & no & no\\
discrete symmetries& $\mathbb Z_2^\text{PC}$ & $\mathbb
Z_2^\text{T}\times\mathbb Z_2^\text{P}\times\mathbb Z_2^\text{C}$ &  $\mathbb
Z_2^\text{TR}$ & $\mathbb Z_2^\text{TPR}\times\mathbb Z_2^\text{PC}$& $\mathbb
Z_2^\text{T}\times\mathbb Z_2^\text{P} \times\mathbb Z_2^\text{R}\times \mathbb Z_2^\text{C}$\\
super\-symmetries & one & none & one & one & none
\end{tabular}
\end{ruledtabular}
\end{table*}

Apart from Lorentz transformation, the continuum model is
(irrespectively of the concrete form of the superpotential)
also invariant under time reversal and parity transformations%
\begin{equation}
\label{eq:wz2dsym8}
\mathbb Z_2^\text{T}\colon (z,\bar z) \mapsto (-\bar z,- z),\quad
 \mathbb Z_2^\text{P}\colon (z,\bar z)\mapsto (\bar z,z).
\end{equation}
Barring possible Wilson terms, the unimproved lattice models obviously
inherit all discrete symmetries from the continuum. By contrast, the
supersymmetrically improved lattice models are invariant only under a
combination of all symmetries. We find 
\begin{subequations}
\label{eq:wz2d:sym10}
\begin{align}
\mathbb Z_2^\text{R}\colon & W'_x(\partial\varphi)_x \mapsto
-W'_x(\partial\varphi)_x,\\
\mathbb Z_2^\text{C}\colon & W'_x(\partial\varphi)_x \mapsto
\overline{W}'_x(\partial\bar\varphi)_x.
\end{align}
\end{subequations}
Thus, for the improved models (with SLAC fermions) the continuum
symmetry is reduced,
\begin{multline}
\label{eq:sym12}
\mathbb Z_2^\text{T}\times \mathbb Z_2^\text{P} 
\times \mathbb Z_2^\text{R} \times \mathbb
Z_2^\text{C}\qquad\longrightarrow\\ \mathbb
Z_2^\text{TPR}\!\times\!\mathbb Z_2^\text{PC}:= \diag(\mathbb Z_2^\text{T}\!\times\!\mathbb Z_2^\text{P}\!\times\!\mathbb Z_2^\text{R})\times
\diag(\mathbb Z_2^\text{P}\!\times\!\mathbb Z_2^\text{C}).
\end{multline}
Here, the diagonal subgroup $\diag(\mathbb Z_2^\text{P}\times\mathbb Z_2^\text{C})$
is a group $\mathbb Z_2^\text{PC}$ generated by the product of the
generators of $\mathbb Z_2^\text{P}$ and $\mathbb Z_2^\text{C}$ (analogous notations
are used for the other groups).
It readily follows that the improvement term must have a 
vanishing expectation value in the original ensemble 
without improvement. We have checked this with a large
numerical precision.
For Wilson and twisted Wilson fermions with improvement
the r.\,h.\,s. of \eqref{eq:sym12} is even further broken down
due to the presence of the (twisted) Wilson term in the
superpotential. For Wilson fermions, the bosonic action
can be read off from \eqref{eq:wz2d:nic11} and \eqref{eq:stdWilson},
\begin{equation}
\label{eq:wz2d:sym13}
\SB=\half\sum_x 
\Big|(\bar\partial\bar\varphi)_x + W'_x
-\textstyle{\frac{r}{2}}(\Delta\varphi)_x\Big|^2.
\end{equation}
Since $\Delta_{xy}$ is invariant under both
time reversal and parity, \eqref{eq:wz2d:sym10} cannot be preserved;
the Wilson term inevitably changes sign. Conversely, from the bosonic
action with twisted Wilson fermions 
\begin{equation}
\label{eq:wz2d:sym13a}
\SB=\half\sum_x 
\Big|(\bar\partial\bar\varphi)_x + W'_x
+\textstyle{\frac{ir}{2}}(\Delta\varphi)_x\Big|^2.
\end{equation}
only $(\varphi\to -m/g-\bar\varphi$,
$\partial\to-\bar\partial$) can be shown to yield a
symmetry. In either case the breaking of the other symmetries is
induced by a higher-dimensional operator and may be expected to
be at most $\ord(a)$ \cite{Catterall:2001fr,Giedt:2005ae}. 
Nevertheless, at finite lattice spacing, the physics might be
affected since the overall size of the breaking terms is a
dynamical question. By contrast, SLAC fermions with the larger symmetry
\eqref{eq:wz2d:sym10} are again favored.

In Tab.~\ref{tab:wz2d:fermions} we summarize all lattice models
to be dealt with in the next section.

\section{Numerical Results}\label{sec:numresults}
\noindent
As outlined in the introduction we have employed the standard
hybrid Monte Carlo algorithm for our numerical simulations.
The fermion determinant was estimated stochastically
utilizing real pseudo-fermion fields. The reason for real
pseudo-fermions derives from the presence of only a single
flavor such that the square root of the pseudofermionic
kernel $Q^{-1}=(MM^T)^{-1}$ is actually needed. We note in
passing that the pseudo fermion action remains real with this
choice since also the fermion matrix is real for Majorana
basis is chosen. Hence the latter was adopted for all our
simulations. A significant gain was achieved by combining
higher order integrators with Fourier acceleration
techniques. With the help of the former one can avoid the
requirement for ever smaller time-step sizes during the MD
step of the HMC while a careful tuning of the latter allows
for autocorrelation times $\tau\leq 5$ over the whole range
of parameters analysed. In particular for small
lattice spacings, i.e.\ at large lattice sizes this was seen
to reduce significantly critical slowing down as also
reported in~\cite{Catterall:2001jg}.
A detailed account of the algorithm employed here will be
published separately at a later time.

\subsection{Dynamical properties of improved lattice
actions}
\begin{figure}
  \input{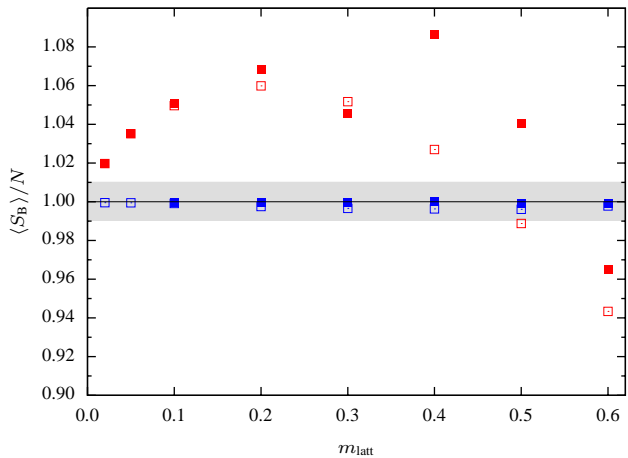}
  \caption{Normalized bosonic action as a
  function of the bare mass lattice parameters using Wilson
  fermions with the improved (filled squares) and 
  unimproved (empty squares) actions from either quenched (red) or dynamical
  fermion (blue) simulations ($N=16\times 16$). }
  \label{fig:wilsonBosonAction}
\end{figure}

\begin{figure}
  \input{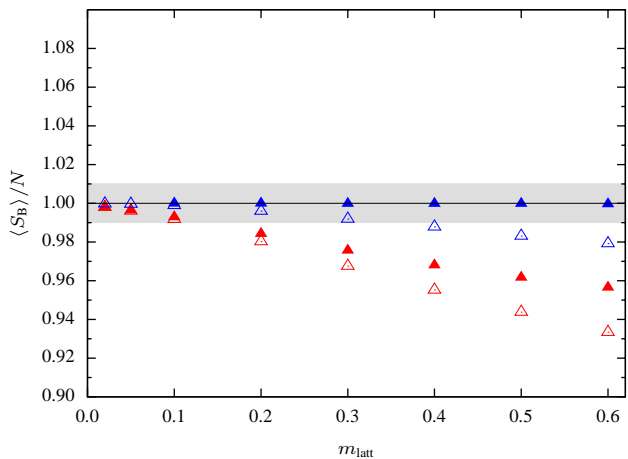}
  \caption{Same as in
  Fig.~\ref{fig:wilsonBosonAction} but for SLAC fermions.}
  \label{fig:slacBosonAction}
\end{figure}

\noindent
Before discussing measurements of physical observables
in the next section we will first focus on the improvement
term~\eqref{eq:wz2d:nic12}. The aim is to understand the difference
between improved and unimproved lattice models w.r.t.\
predictions of supersymmetry.
One possible test is a measurement of the bosonic action
itself. With the help of the Nicolai map appearing in~%
\eqref{eq:wz2d:nic6} one can show that
\begin{equation}
\label{eq:impr1}
\langle \SB \rangle = N. 
\end{equation}
Here, $N=\Nt\times \Ns$ denotes the total number of lattice
points, and \eqref{eq:impr1} is only expected to hold when fermions are
included dynamically. Then, however, this prediction holds
irrespectively of the concrete value of the coupling
constants. With a slightly different argument
the same was also found in~\cite{Catterall:2001fr}.  Equation \eqref{eq:impr1}
provides a test observable distinguishing improved from
unimproved lattice models as well as quenched from
dynamical fermion simulations. To accomplish this, we have 
run simulations with both (standard) Wilson and SLAC
fermions. The results are shown as a function of the bare
lattice mass parameter $\mlatt=m/\Ns$. Since the continuum
limit for this theory is obtained from $\mlatt\to 0$,
smaller values of $\mlatt$ likewise mean a finer lattice
spacing (and for fixed $N$ a smaller spacetime volume).
The dimensionless coupling strength $\lambda=g/m$ was set
to $\lambda=1$. The lattice sizes we used for our numerical
simulations were $N=16\times 16$ for Wilson and
$N=15\times 15$ for SLAC fermions. For the quenched simulations
500,000 (independent) configurations were evaluated, and
30,000 configurations with dynamical fermions were analysed.
The results are shown in Fig.~\ref{fig:wilsonBosonAction}
for Wilson and in Fig.~\ref{fig:slacBosonAction} for SLAC
fermions.
One clearly observes that the quenched data significantly deviate from
the predicted value which illustrates the
necessity of dynamical fermion contributions in order to retain 
supersymmetry. Using an unimproved action with 
dynamical fermions we find much smaller
deviations which in case of the Wilson fermions are
already hard to distinguish from the improved results. For
SLAC fermions the deviations are somewhat more systematic and
remain also clearly distinguishable from other dynamical
fermion simulations. A second difference between Wilson and
SLAC fermions may be infered from
\begin{figure}
  \input{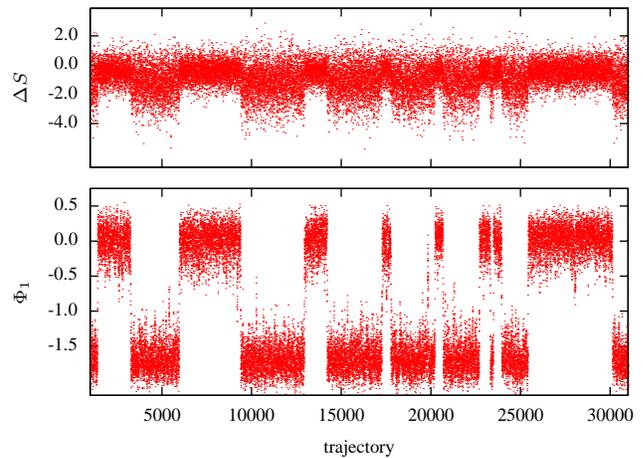}
  \caption{MC history of the lattice mean 
  $\Phi_1=N^{-1}\sum_x \varphi_{1,x}$ and size of the
  improvement term for Wilson fermions ($N=16\times 16, \lambda=0.6, \mlatt=0.3$).}
  \label{fig:couplingWilsonImprovement}
\end{figure}
Fig.~\ref{fig:couplingWilsonImprovement}. Namely, there is
a distinct correlation between the ground state around which
the field $\varphi_1$ fluctuates on the one hand and
size and variance of the improvement term on the
other hand. This may be taken as  direct manifestation
of the additionally broken $\mathbb Z_2^\text{TPR}$-symmetry due to the
Wilson term which will also play a role when discussing
the failure of improvement in the next paragraphs.

\subsubsection*{Limitations of improved lattice actions}
\noindent
Studying the improvement term $\Delta S$ for models with
either Wilson or SLAC fermions we have found that the
system is ultimately pushed into an unphysical region of
configuration space, at least for strong coupling. Our
simulations have revealed that this instability is
controlled by the actual size of the bare mass parameter
and the coupling strength $\lambda$. Simulations tend to
fail more often as either of them grows. The study of
this phenomenon with Wilson fermions turns out to be clumsy
since there is no clear correlation between the value of the
coupling and the number of configurations where the instability
occurs. Hence we prefer to present our analysis from the
simulations with SLAC fermions. However, it should be emphasized
again that for either Wilson or twisted Wilson fermions the
qualitative picture is the same as described below.

\begin{figure}
   \input{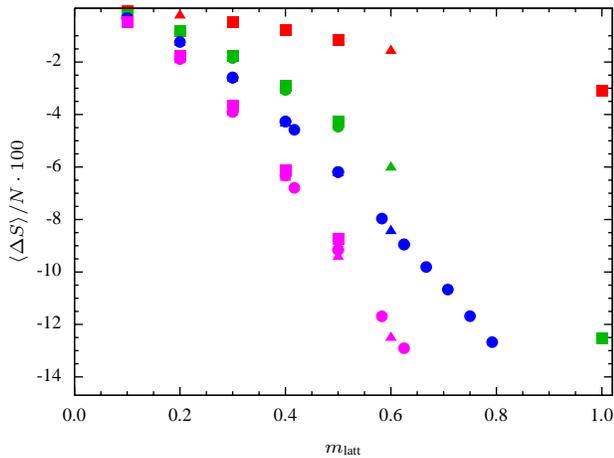}
  \caption{Reduced improvement term $\Delta
  S/N$ for different lattice sizes: $9\times 9$
  (squares), $15\times 15$ (triangles) and $25\times 25$ (circles). 
  Colors depict $\lambda=$ 0.8 (red), 1.0 (green), 1.2
  (blue),  1.5 (magenta).}
  \label{fig:improvementTermSlac}
\end{figure}

\begin{figure}
   \input{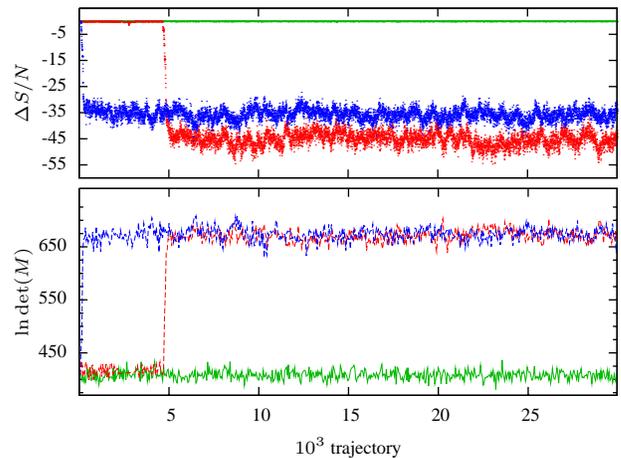}
  \caption{MC history of improvement term and
  fermion determinant (SLAC improved, $N=15\times
  15$, $\mlatt=0.6$, $\lambda=$ 1.4 (green), 1.7 (red),
  1.9 (blue)).}
  \label{fig:crashedSlac8}
\end{figure}

\begin{figure}
   \input{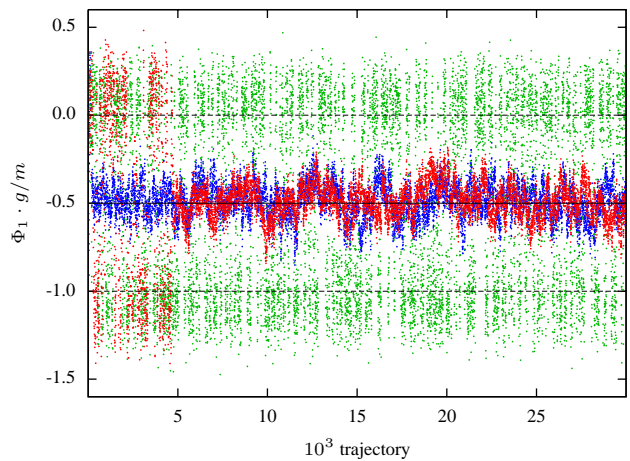}
  \caption{MC history of normalized lattice mean $\Phi_1\!\cdot\!g/m$
  (SLAC~impr., $N=15\times
  15$, $\mlatt=0.6$, $\lambda=$ 1.4 (green), 1.7 (red), 1.9
  (blue)).}
  \label{fig:crashedSlac9}
\end{figure}

\begin{figure}
  \input{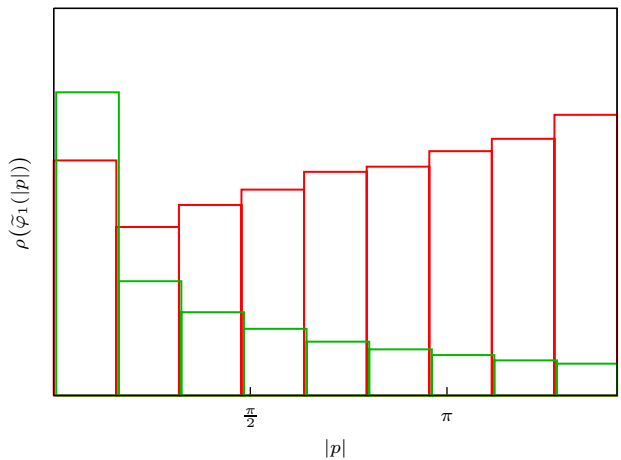}
  \caption{Mode analysis of ensembles in the
  physical (green, $\lambda=1.4$) and unphysical (red, 
  $\lambda=1.7$) phase. Here $\rho$ is the distribution function
  for the modulus of the lattice momentum averaged over 25,000 configurations
   (SLAC improved, $N=15\times 15$, $\mlatt=0.6$).}
  \label{fig:mode_analysis}
\end{figure}

It is to be expected that the improvement term grows with the
coupling strength $\lambda$ and vanishes continuously
in the continuum limit (at $\mlatt=0$). We observe a good
scaling behavior w.r.t.\ the lattice size,
see also Fig.~\ref{fig:improvementTermSlac}. For all
couplings $\lambda$ and $\mlatt$ the improvement term is 
found to be smaller than 14\% of the total bosonic action.
Depending on the coupling strength $\lambda$, this ratio is
reached sooner or later. Actually, this represents a
threshold above which the simulation fails. The situation is
depicted in Figs.~\ref{fig:crashedSlac8} and~\ref{fig:crashedSlac9}.
At some instant, the improvement term blows up and settles again
at a value about 40 times the size of the bosonic action.
At the same time also the fermion determinant grows drastically
and so hinders the system from returning into the original
(and desired) region of configuration space. A reason for this
instability may be found by reconsidering the improved action 
\begin{equation}
\label{eq:impr2}
\SB=\frac{1}{2}\sum_x \Big| 2(\partial\varphi)_x +
\overline{W}_x \Big|^2.
\end{equation}
In this form the action allows for two distinct behaviors
of the fluctuating fields. The physically expected behavior consists
of small fluctuations around the classical minima of the potential.
Alternatively, \eqref{eq:impr2} allows for large fluctuations of
$\varphi$ to be compensated by large values of $\overline{W}_x$.
The latter would be dominated by UV contributions, and this is what we
actually observe, cf.~Fig.~\ref{fig:mode_analysis}. 
In this situation, it is definitely no longer possible to
extract meaningful physics. Another view on this ``broken'' phase is
taken in Fig.~\ref{fig:crashedSlac9}. While the
ensemble with $\lambda=1.4$ exhibits the expected behavior
at the only slightly larger value of $\lambda=1.7$ the
simulation breaks down after about 5,000 configurations and
for $\lambda=1.9$ the simulation is instantly found in the
broken phase. 

To sum up, we have observed that the improved lattice
models may become unstable at any finite $\mlatt$ and hence any finite
lattice spacing. If and when this happens depends on several factors.
Wilson fermions are affected in a stronger way while SLAC
fermions remain stable for a much wider range of coupling constants.
Apart from that, one should ensure by monitoring the
improvement term or any other observable discussed above explicitly
that a simulation is not subject to this phenomenon. For the
practitioner this is of course a major nuisance and possibilities
to avoid this matter are already under investigation. 
Provided that one is confined to lattices smaller than
$64\times 64$ but demands the absence of finite-size
effects, improved lattice models with Wilson
fermions can be used for the continuum extrapolation of masses
only up to $\lambda<0.4$. 
SLAC fermions can be used in the greater range of $\lambda<1.5$;
the corresponding results will be presented further below.

\subsection{Setting the stage}
\noindent
In Monte-Carlo simulations, importance sampling
is only meaningful  with respect to a positive measure.
However, including dynamical fermions the measure is
$\det(M)\exp(-\SB)$. While the exponential factor is strictly
positive ($\SB$ is real), the positivity of the determinant
cannot be guaranteed for an interacting theory and a
possibly emerging sign problem has to be
addressed. In order to make sensible comparisons with
continuum calculations (which are most conveniently performed in an
infinite spacetime) one furthermore must make sure that
physical observables extracted from lattice simulations are
free of finite-size effects. In order to check this, all
simulations in this section are repeated in portions of
fractional volume $l^2$ of a fixed physical unit volume
(with various values for $l=\Ns a$ on a square lattice with
$N=\Nt\times\Ns$ lattice points).
In the following we consider both issues in more detail.

\subsubsection{Negative fermion determinants}
\noindent
The Nicolai map in a supersymmetric theory is a change of bosonic variables
which renders the bosonic part of the action Gaussian; at the same time, the
Jacobian of this change of variables has to cancel the fermion determinant. In
our model, this means
\begin{equation}
\det(M) = \det\left(\frac{\delta}{\delta
\varphi}\big( 2(\bar\partial \bar\varphi) + W' \big) \right).
\end{equation}
In this light, an indefinite fermion determinant obviously corresponds to
a non-invertible change of variables in the continuum,
\begin{equation}
\varphi\mapsto \xi = 2\bar\partial \bar\varphi + W'.
\end{equation}
This map is only globally invertible if the superpotential is of degree~1
(the Nicolai map in this case has winding number~1), i.e., for the free
theory~\cite{Cecotti:1981fu}. For our choice $W'(\varphi) = m\varphi + g\varphi^2$
the map is not \emph{globally} invertible, and there exists at least one
point where $\det(M)$ vanishes iff $g\neq 0$. By this line of argument
(for the continuum formulation of the model) negative determinants
cannot be ruled out.

\begin{figure}
\input{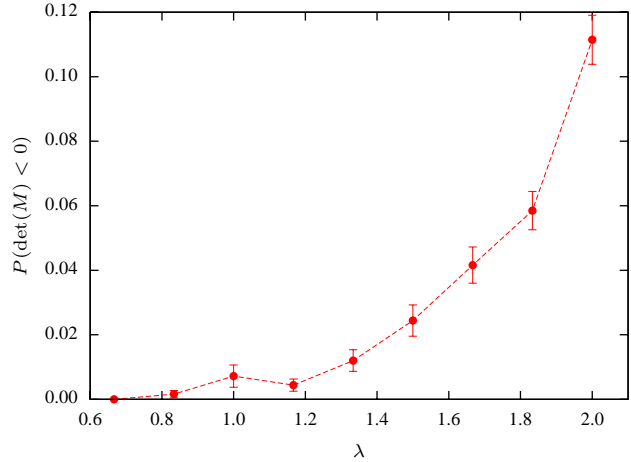}
\caption{\label{fig:detOverLambda}Probability for negative
determinants (Wilson unimproved, $N=14\times 14$,
$\mlatt=0.43$).}
\end{figure} 
\begin{figure}
\input{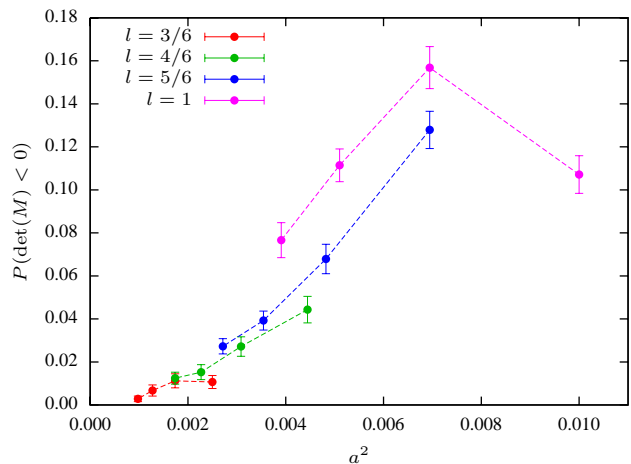}
\caption{\label{fig:determinantOverA}Probability for negative determinants at
different box sizes with varying lattice size (Wilson
unimproved, $m=6$, $\lambda=2.0$).}
\end{figure}
One way to cope with this in practical simulations is to use
$\abs{\det(M)}\exp(-\SB)$ for the generation of configurations
instead and to reweigh with the sign afterwards. Unfortunately,
calculating the sign of $\det M$ is as costly as the computation
of the whole determinant. Hence, this method becomes unfeasible
for large lattices. A way out is to avoid reweighing within
certain bounds for the parameters in which the ensuing systematic errors
are negligible. Thus, we have to estimate the frequency of occurrence
of negative determinants as a function of the parameters.
To obtain more reliable results we have studied this
subject with a naive inversion algorithm which computes the
determinant from a LU decomposition and takes its 
contributions exactly into account. This is numerically
much more involved than the standard pseudo-fermion
algorithm, thus, this method is only applicable to small lattice
sizes with up to $16\times 16$ lattice sites.
For fixed physical mass $m$ it can be gleaned from
Fig.~\ref{fig:detOverLambda} that configurations with a negative
sign of the determinant show up only for $\lambda>1.0$.
Furthermore, in order to understand the dependence on the lattice
size and the lattice spacing we have fixed the coupling to
$\lambda=2.0$ and run simulations on fractions $l^2$ of a unit
physical volume ($l\in\{3/6,4/6,5/6,6/6\}$) and
different lattice spacings. The results displayed in
Fig.~\ref{fig:determinantOverA} clearly show that the
problem dissolves in the continuum limit but becomes worse at
every finite lattice spacing when the physical volume is
increased. For both figures, for each data point about 50,000
configurations were evaluated.
Eventually, to estimate the impact on actual measurements
we have measured the bosonic action with $m=5$, $\lambda=2$ on a
$12\times 12$ lattice and obtained about $7\% $
configurations with a negative sign for the fermion
determinant. The expectation values considered here
are $\vev{\SB}_\text{non-reweighed} =
149.94(12)$ and $\vev{\SB}_\text{reweighed} = 149.49(10)$.
Hence even at large coupling (far larger than
what we target at in the next section) effects may be assumed to
be at most of marginal relevance for actual measurements. 

\subsubsection{Finite size effects}
\begin{figure}
\input{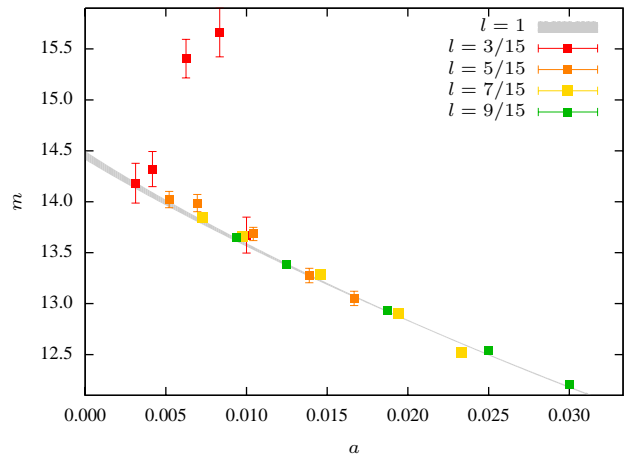}
\caption{\label{fig:finiteSizeWilsonImp}Lattice masses for $m=15$,
$\lambda=0.3$ on box sizes $l\in[0.2,0.6]$. We see a systematic deviation from the
$l=1$ result below $l\approx 0.5$.}
\end{figure}

\noindent
For these models the bare mass $\mlatt$ also sets
the scale for the overall spacetime volume. As with all
lattice simulations we have to balance finite-size and
discretization errors. If the lattice spacing is
chosen too large, lattice artifacts may
grow; on the other hand if, say, the Compton
wavelength of a particle is larger than the spacetime volume
the extraction of masses may suffer from finite-size effects.
One way to test for the presence of such finite-size
violations is to study the model at different spacetime
volumes. Comparing the fermion species introduced earlier
Wilson fermions may be expected to be most affected. Here,
lattice artifacts further increase the correlation lengths
so that measurements are much more sensitive to the finite
box size. Our setup for this analysis is as follows.
At first we have simulated the improved lattice model using
Wilson fermions at fixed coupling parameters $m=15$ and
$\lambda=0.3$ for five different lattices with $\Nt=\Ns\in\{20, 24,
32, 48, 64\}$ lattice points in each direction ($N=\Nt\times\Ns$).
In the following we assume that with this choice of coupling
constants the spacetime volume is large enough so as to allow
for a sufficiently good identification with the thermodynamic
limit. The masses obtained from these simulations
were extrapolated to the continuum as described in
App.~\ref{app:contExtr}. This is also shown in
Fig.~\ref{fig:finiteSizeWilsonImp} where the resulting fit
(and its uncertainty) is depicted with a gray shaded area.
The next step is to decrease the volume to fractions
$l^2$ (with $l \in \{9/15,7/15,5/15,3/15\}$) of a fixed
physical unit volume. As long as no finite-size effects are
visible we expect to find the masses extracted at these smaller
and smaller volumes to lie on top of the fit from the original
lattice (of unit volume). Up to a volume less than
half the size of the original one this scaling may be
easily infered from Fig.~\ref{fig:finiteSizeWilsonImp}
which justifies a posteriori the correctness of our earlier
assumption.

However, since by perturbation theory the physical masses
decrease for growing coupling (see next section), we expect
growing Compton wavelengths and therefore stick to unit volume
($l=1$) for all further measurements so as to exclude finite-size
effects.

\subsection{Weak coupling}
\label{ssec:weakCoupling}
\noindent
An interesting observable for comparing lattice results with continuum
physics is the mass of the lightest excited state, i.e.\ the energy gap.
Since unbroken supersymmetry in the continuum predicts that bosonic and
fermionic masses coincide it also provides a possibility 
to check the supersymmetric properties of the lattice prescription.
The corresponding values can be extrapolated from the lattice masses in the
continuum limit. In the weak coupling it will be possible to match these results
to predictions of perturbation theory. This provides an important test for the
numerical results and ensures that also the results at intermediate
coupling are reliable.

For a description of our prescription for the boson and fermion mass
extraction from correlators on the lattice we refer the interested reader
to App.~\ref{app:masses}. With these methods we are aiming at a test of the lattice
results against perturbation theory for $\lambda\le 0.3$.

The reference value is given by a one-loop calculation of the
renormalized mass 
\begin{equation}
\label{eq:oneLoopResult}
m_\text{ren}^2  = m \left(1-\frac{4\lambda^2}{3\sqrt{3}}\right) +\ord(\lambda^4)
\end{equation}
in the continuum valid for $\lambda\ll 1$ with the bare mass $m$
as used in Eq.~\eqref{eq:superpot}. To obtain this result one first
must calculate contributions of the loop diagrams to the propagator.
An expansion in $\lambda$ then yields the above result.\footnote{We will
elaborate on the analytical side and the determination of the effective
potential of this theory in a forthcoming publication.}

As will be show below the fermionic masses have lower statistical errors than
the bosonic ones. Therefore we compare only the extrapolations
for fermionic masses to the perturbative results. This procedure gets justified
by the fact that bosonic and fermionic masses coincide even on a finite lattice
for the weak coupling regime as described below in Sec.~\ref{sssec:bosonsFermions}.

\subsubsection{Continuum limit}
\noindent
The methods to extrapolate to the continuum given in App.~\ref{app:contExtr} are
based on the free theory with $\lambda=0$. Since we are interested in the
interacting case we must first make sure that the continuum extrapolation of masses
remains stable even for $\lambda=0.3$.

To that purpose we consider the masses in the improved model with
standard Wilson and twisted Wilson fermions at $\lambda=0.3$ at different
lattice spacings $a$. In the perturbative coupling regime we use throughout
square lattices of sizes $\Nt=\Ns
\in \{ 20, 24, 32, 48, 64\}$. These correspond to lattice spacings of about
$a\in [0.015625,0.05]$. A statistics of 10,000 independent configurations
puts us in a position to extrapolate to the continuum.

\begin{figure}
   \input{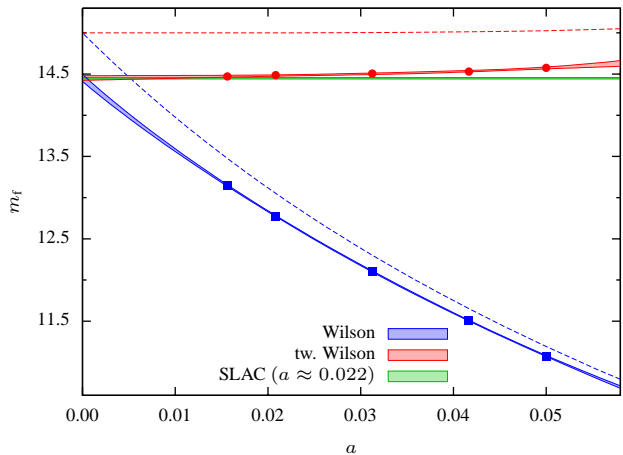}
\caption{\label{fig:interactingContinuum}
The continuum extrapolation of masses for $\lambda=0.3$ for the improved Wilson
and twisted Wilson model. Here, the SLAC result is given for one single lattice size.
For comparison the exact results for the free theory are also shown.}
\end{figure}

Using these masses $m(a)$ at finite lattice spacing the extrapolation is shown
in Fig.~\ref{fig:interactingContinuum}. For comparison we also mark the
mass for SLAC fermions at a finite lattice size $\Nt=\Ns=45$
(corresponding to $a\approx 0.022$). All these results indicate that even at
$\lambda=0.3$ the continuum extrapolated masses coincide within
error bounds. Even better, the masses of SLAC fermions at finite
lattice spacing can not be distinguished from the continuum result.

\subsubsection{Comparison with perturbation theory}
\begin{figure}
  \input{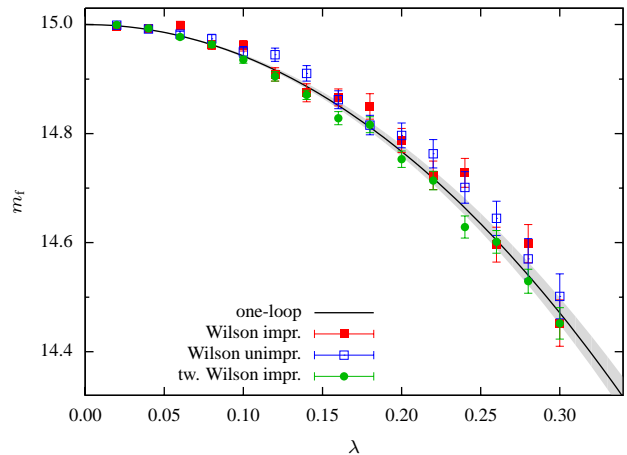}
\caption{\label{fig:perturbativeMasses}Continuum masses for the weakly coupled
regime in comparison to the perturbative result. The shaded area corresponds to
the extrapolation provided by the continuum results according to
Eq.~\eqref{eq:perturbativeParameters} with $m=15$ and $b=1.35(13)$.}
\end{figure}

\noindent
As described above we extrapolate masses for
Wilson (improved and unimproved) and twisted Wilson (improved) fermions for $\lambda\in[0,0.3]$
to the continuum values, cf.
Fig.~\ref{fig:perturbativeMasses}.%
\begin{table}
\caption{\label{tab:weakCoupling}Continuum extrapolations of fermionic masses
for Wilson and twisted Wilson fermions in the weak coupling regime.}
\begin{ruledtabular}
\begin{tabular}{lccc}
$\lambda$ & Wilson unimp. & Wilson imp. & tw. Wilson \\ \hline
$0.02$ & $14.999(2)$ & $14.997(2)$ & $14.999(1)$ \\
$0.04$ & $14.992(4)$ & $14.993(4)$ & $14.993(3)$ \\
$0.06$ & $14.982(6)$ & $14.999(7)$ & $14.977(4)$ \\
$0.08$ & $14.974(8)$ & $14.963(8)$ & $14.963(5)$ \\
$0.10$ & $14.95(1)$ & $14.96(1)$ & $14.935(6)$ \\
$0.12$ & $14.94(1)$ & $14.91(1)$ & $14.905(9)$ \\
$0.14$ & $14.91(1)$ & $14.87(2)$ & $14.871(9)$ \\
$0.16$ & $14.86(2)$ & $14.87(2)$ & $14.83(1)$ \\
$0.18$ & $14.82(2)$ & $14.85(2)$ & $14.82(1)$ \\
$0.20$ & $14.80(2)$ & $14.79(2)$ & $14.75(2)$ \\
$0.22$ & $14.76(3)$ & $14.72(3)$ & $14.71(2)$ \\
$0.24$ & $14.70(3)$ & $14.73(3)$ & $14.63(2)$ \\
$0.26$ & $14.64(3)$ & $14.60(3)$ & $14.60(2)$ \\
$0.28$ & $14.57(4)$ & $14.60(4)$ & $14.53(2)$ \\
$0.30$ & $14.50(4)$ & $14.45(4)$ & $14.45(3)$
\end{tabular}
\end{ruledtabular}
\end{table}
The masses coincide within error bars although the twisted Wilson
masses are systematically smaller. This difference has to be interpreted as a
systematic error in the continuum extrapolation for the masses but its effect is
almost overshadowed by the statistical errors in our case. However this result 
indicates that for a reliable extrapolation at larger statistics finer lattices
can be necessary to yield a better continuum limit.

As a further test we use these results to reproduce the perturbative formula
\begin{equation}
\label{eq:perturbativeParameters}
m(\lambda) \approx m_0\sqrt{1-\frac{\lambda^2}{b}}.
\end{equation}
Taken this functional form for granted, the parameters $m_0$ and $b$ can be extracted
from a least-square fit to the given data. For this fit we can use our knowledge
about the free theory ($m_0=15$) as a fixed input or, alternatively, allow for
both $m_0$ and $b$ as free parameters. The corresponding results
are given in Tab.~\ref{tab:perturbativeMass}. 

\begin{table}
\caption{\label{tab:perturbativeMass}Fit for the perturbative mass formula with 
$\ord(\lambda^2)$ corrections to be compared with the one-loop results. For
comparison the one-loop result is $b\approx 1.2990$.}
\begin{ruledtabular}
\begin{tabular}{lcc}
derivative & $b$ & $m_0$\\ \hline
Wilson improved & $1.34(6)$ & $15.007(6)$ \\
Wilson unimproved & $1.39(7)$ & $15.008(6)$ \\
twisted Wilson improved & $1.26(4)$ & $14.996(4)$ \\ \hline 
Wilson improved & $1.37(5)$ & fixed to $15$ \\
Wilson unimproved & $1.42(6)$ & fixed to $15$ \\
twisted Wilson improved & $1.25(3)$ & fixed to $15$ 
\end{tabular}
\end{ruledtabular}
\end{table}

The extrapolated results for $m_0$ confirm that the extrapolation to the free
theory works reliably and that we can expect to obtain meaningful results for
$b$. Furthermore the results obtained for improved and unimproved Wilson
fermions coincide very well and therefore both provide the correct continuum
limit.

Additionally the results for standard Wilson and twisted Wilson fermions lead to
compatible results when taking systematic uncertainties of the continuum
extrapolation into account.

As an important result of these observations, all three models considered in
the weak coupling case tend towards the same continuum limit for $\lambda>0$. The
perturbative results can be recovered where the largest error bars (including
possible systematic errors) yield $b=1.35(13)$ in agreement with the one-loop
result of $b_\text{one-loop}\approx 1.2990$.

\subsubsection{Signs of supersymmetry at finite lattice spacing}
\label{sssec:bosonsFermions}
\noindent
Apart from all results solely based on fermions, we are primarily interested
in the restoration of supersymmetry on the lattice. For this reason we better also
check the demand from supersymmetry that the masses of bosonic and
fermionic superpartners match. This is going to be checked by computing bosonic
and fermionic masses at couplings $\lambda=0.2$ and $\lambda=0.4$ with $m=15$
for all the models on different lattice sizes.

As we have seen in the whole weak coupling regime the fermionic masses do not
suffer from statistical noise. This behavior derives from the fact that the
fermionic correlator for the free theory ($\lambda=0$) is independent of the
bosonic field $\varphi$ and is obtained by a pure matrix inversion. At small
(and finite) $\lambda$, corrections to the free propagator are of
$\ord(\lambda^2)$, and the fluctuations of $\varphi$ during the simulation
are suppressed with $\lambda^2$; a statistics of only $10^4$ is needed to
get reliable results.

On the other hand the bosonic correlator even for the free theory is given by
the correlations of the fluctuating field $\varphi$. Therefore a much higher
statistics is necessary to sample the bosonic two-point function. Here, problems
arise by the exponentially growing relative error of the two-point function
$C(t)$ with respect to $t$.

Only with the use of an algorithm combining Fourier acceleration with
higher order integrators it was possible to simulate $10^6$ to $10^7$
configurations for each parameter set $(m,\lambda)$ with an autocorrelation
time of the two-point function of $\tau\le 2$.

The results of these numerical efforts are summarized in
Tab.~\ref{tab:massesBosonFermion}. They show that independently of the
model even for $\lambda\in\{0.2,0.4\}$ bosonic and fermionic masses correspond
to each other and lattice-induced supersymmetry breaking can not be observed.

\begin{figure}
   \input{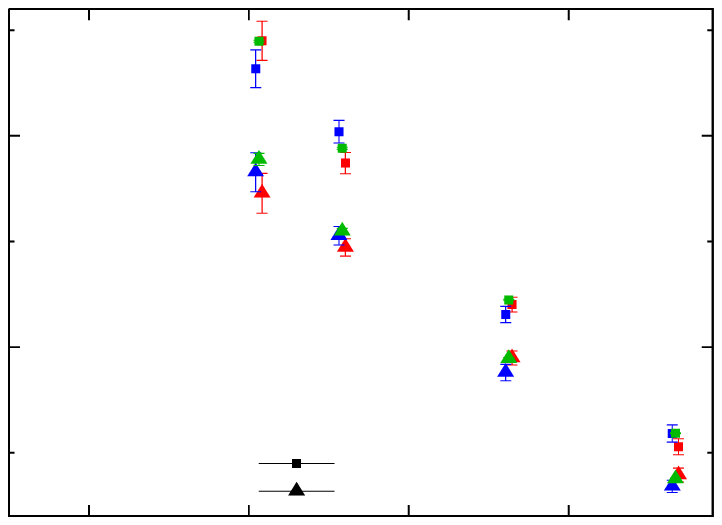}
\caption{\label{fig:bosonMassesWimp}Bosonic and fermionic masses for the weakly
coupled regime for the improved Wilson model.}
\end{figure}

\begin{figure}
   \input{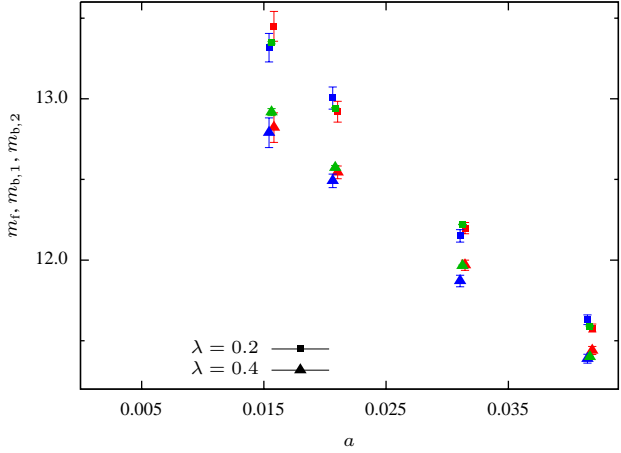}
\caption{\label{fig:bosonMassesWuimp}Bosonic and fermionic masses for the weakly
coupled regime for the unimproved Wilson model.}
\end{figure}

Finally in Figs.~\ref{fig:bosonMassesWimp} and~\ref{fig:bosonMassesWuimp} the 
derived bosonic and fermionic masses are shown
for the improved (and unimproved) model with Wilson fermions. Even these high
statistics do not allow for a clear cut distinction between the extrapolated
continuum masses of bosons and fermions for the improved and the
unimproved models. This proves that even at $\lambda=0.4$ the improvement is
not necessary even on a finite lattice. Each model tends towards the
supersymmetric continuum limit.

\begin{table}
\caption{\label{tab:massesBosonFermion} For different models and lattice sizes
we computed bosonic and fermionic masses with bare mass $m=15$.}
\begin{ruledtabular}
\begin{tabular}{lccccc}
model & $\Ns$ & $\lambda$ & $m_\text{f}$ & $m_{\text{b},1}$& $m_{\text{b},2}$ \\ \hline
Wilson impr. & $24$ & $0.2$ & $11.592(2)$ & $11.53(4)$ & $11.59(4)$ \\
& $24$ & $0.4$ & $11.375(4)$ & $11.39(3)$ & $11.34(3)$ \\
& $32$ & $0.2$ & $12.224(2)$ & $12.20(3)$ & $12.15(4)$ \\
& $32$ & $0.4$ & $11.945(5)$ & $11.95(3)$ & $11.88(4)$ \\
& $48$ & $0.2$ & $12.941(5)$ & $12.87(5)$ & $13.02(5)$ \\
& $48$ & $0.4$ & $12.548(13)$ & $12.47(4)$ & $12.53(4)$ \\
& $64$ & $0.2$ & $13.349(10)$ & $13.45(9)$ & $13.32(9)$ \\
& $64$ & $0.4$ & $12.89(3)$ & $12.73(9)$ & $12.83(9)$ \\ \hline
Wilson unimpr. & $24$ & $0.2$ & $11.591(2)$ & $11.58(2)$ & $11.63(3)$ \\
& $24$ & $0.4$ & $11.400(4)$ & $11.44(2)$ & $11.39(3)$ \\
& $32$ & $0.2$ & $12.221(2)$ & $12.20(3)$ & $12.15(4)$ \\
& $32$ & $0.4$ & $11.965(5)$ & $11.97(3)$ & $11.87(4)$ \\
& $48$ & $0.2$ & $12.942(5)$ & $12.92(6)$ & $13.00(7)$ \\
& $48$ & $0.4$ & $12.572(14)$ & $12.54(4)$ & $12.49(4)$ \\
& $64$ & $0.2$ & $13.347(7)$ & $13.45(9)$ & $13.32(9)$ \\
& $64$ & $0.4$ & $12.91(2)$ & $12.82(9)$ & $12.79(9)$ \\ \hline
tw. Wilson (impr.)& $24$ & $0.2$ & $14.811(7)$ & $14.94(11)$ & $14.91(12)$ \\
& $24$ & $0.4$ & $14.13(1)$ & $14.21(9)$ & $14.06(8)$ \\
& $32$ & $0.2$ & $14.788(6)$ & $14.61(14)$ & $14.94(12)$ \\
& $32$ & $0.4$ & $14.08(1)$ & $14.39(14)$ & $13.68(13)$ \\
& $48$ & $0.2$ & $14.789(6)$ & $14.74(11)$ & $14.61(11)$ \\
& $48$ & $0.4$ & $14.04(1)$ & $14.16(16)$ & $13.98(15)$ \\ \hline
SLAC impr. & $45$ & $0.2$ & $14.768(4)$ & $14.87(10)$ & $14.83(9)$ \\
& $45$ & $0.4$ & $13.997(13)$ & $14.08(11)$ & $13.92(10)$ \\ \hline
SLAC unimpr. & $45$ & $0.2$ & $14.769(4)$ & $14.75(6)$ & $14.57(6)$ \\
& $45$ & $0.4$ & $14.047(16)$ & $13.74(8)$ & $13.75(7)$ 
\end{tabular}
\end{ruledtabular}
\end{table}

\subsection{Intermediate coupling results}
\noindent Earlier attempts to go beyond the perturbative
regime could not reliably determine the mass spectrum.
Namely, this was hindered by instabilities introduced by
improvement terms. For Wilson fermions, this renders
simulations at intermediate couplings invalid. For our
analysis of coupling constants in the intermediate regime
($0.3<\lambda\le 1.2$) we have therefore only considered
actions with twisted Wilson and SLAC fermions (which are
anyhow expected to give better results at finite lattice
spacing). For twisted Wilson fermions we have run simulations
with the improved action on lattices with
$\Ns\in\{32,40,48,56,64\}$ lattice points in the spatial
direction. For the temporal direction we have used
$1.25\cdot\Ns$ lattice points in order to be able to assess
whether contributions from higher excited states are really
absent. At the chosen value of $m=15$ in all simulations,
the respective bare lattice mass parameter $\mlatt$
confines the attainable coupling strengths to $\lambda\le 0.7$.%
\footnote{For $\lambda=0.7$ we already observed
that the simulation failed on the coarsest lattice and had
to be excluded.}
For even larger coupling strengths $\lambda$ only SLAC
fermions have been found to yield sensible results. In our
simulations we used for this species both the improved and
unimproved lattice models on a fixed lattice size of $N=45\times
45$. Apart from that, one further run was done on a $63\times 63$
lattice with $\lambda=0.8$. Square lattices turned
out to be more convenient with SLAC fermions and to be
sufficient to clearly read off (within statistical errors) the
masses. As for the simulations with twisted Wilson fermions
we have determined only the masses from the fermionic
correlators since with the statistics (50,000
trajectories) achieved so far the bosonic correlators 
are far too noisy to yield reliable results.
\begin{table}
\caption{\label{tab:intermediate}Fermionic masses for the intermediate coupling
case. Twisted Wilson results are continuum extrapolations
whereas the SLAC data is from a $45 \times 45$ lattice.}
\begin{ruledtabular}
\begin{tabular}{lccc}
$\lambda$ & tw. Wilson & SLAC unimp. & SLAC imp. \\ \hline
$0.20$ & $14.80(2)$ & $14.769(4)$ & $14.768(4)$ \\
$0.35$ & $14.23(2)$ &  &  \\
$0.40$ & $13.99(3)$ & $14.05(2)$ & $14.00(1)$ \\
$0.45$ & $13.62(5)$ &  &  \\
$0.50$ & $13.30(6)$ &  &  \\
$0.55$ & $12.8(1)$ &  &  \\
$0.60$ & $12.2(1)$ & $12.81(4)$ & $12.44(6)$ \\
$0.65$ & $11.9(2)$ &  &  \\
$0.70$ & $10.4(5)$ &  &  \\
$0.80$ &  & $11.49(9)$ & $10.2(3)$ \\
$1.00$ &  & $10.2(2)$ & $9.4(2)$ \\
$1.20$ &  & $10.1(3)$ & $9.1(3)$
\end{tabular}
\end{ruledtabular}
\end{table}
\begin{figure}[b]
  \input{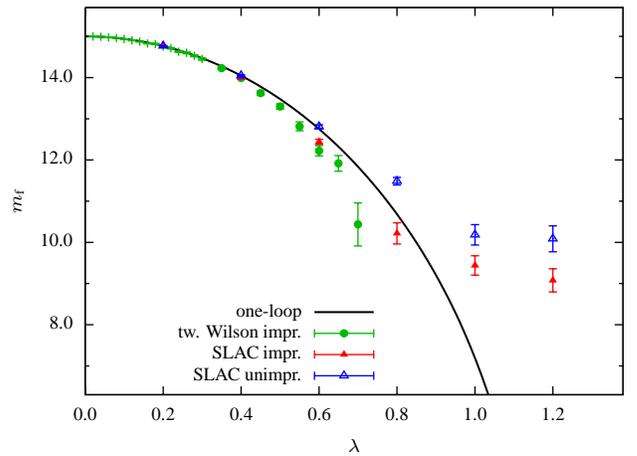}
  \caption{\label{fig:slacStrong}Masses of the improved and unimproved model
  with SLAC fermions on a $45\times 45$ lattice and continuum extrapolated results
  for
  twisted Wilson fermions are compared with the perturbative one-loop result in
  the continuum.}
\end{figure}

Our results may be found in Tab.~\ref{tab:intermediate}
and are depicted graphically in Fig.~\ref{fig:slacStrong}.
From the comparison with perturbation theory first
deviations are seen as soon as $\lambda\geq 0.4$ where
the (extrapolated) lattice results are slightly stronger curved.
Also clear deviations between the improved and unimproved
model using SLAC fermions become apparent for
$\lambda\geq0.6$. It is worthwhile to note that the
result from the improved lattice model is closer to the
continuum limit which may be infered from
Tab.~\ref{tab:slacContinuum}. While the  
lattice data from the improved model almost coincide for
both lattice spacings the data from the unimproved model
is likely to approach the same value if increasingly finer
grained lattices are used.

Larger values of $\lambda$ are attainable however the
numerical effort becomes more involved and some
technicalities need to be addressed. Once this is under
control we are confident to investigate the strong coupling
regime with the improved models up to $\lambda=2.0$ on the
same lattice sizes. The interesting question whether the
masses of superpartners still agree can then be 
satisfactorily answered.
\begin{table}
\caption{\label{tab:slacContinuum}Fermionic masses for the SLAC derivative on
two different lattice sizes for $\lambda=0.8$.}
\begin{ruledtabular}
\begin{tabular}{lcc}
$\Ns$ & improved & unimproved \\ \hline
$45$ & $10.22(26)$ & $11.49(9)$ \\
$63$ & $10.54(15)$ & $10.70(19)$
\end{tabular}
\end{ruledtabular}
\end{table}

\section{Conclusions and outlook}
\noindent
In this article we have presented  a detailed
numerical analysis of the two-dimensional $\mathcal N=(2,2)$
Wess-Zumino model. Due to algorithmic improvements we were
able to study lattice models at much larger lattice sizes,
i.e.\ smaller lattice spacings and more importantly at stronger
couplings. For a comparison with analytical results
from perturbation theory we have checked explicitly for
finite-size effects and other systematic
errors such as sign changes of the fermion determinant.
Both were seen to be under control for the scrutinized
parameter range. We could confirm earlier weak
coupling results and for the first time resolve deviations
from perturbation theory. All three kinds of fermions,
Wilson, twisted Wilson, and SLAC fermions, approach the same
continuum results. It turned out that lattice artifacts were
largest for Wilson and smallest for SLAC fermions. At intermediate
coupling we observed that the supersymmetrically improved lattice
action using Wilson fermions lead to unstable simulations
that eventually fail to produce reliable results unless very
large lattices are chosen. Simulations with SLAC fermions proved
to be much more stable; they allow for improvement terms
for a wider parameter range. At finite lattice spacing and
weak coupling no significant differences in the measured spectrum
between simulations using the improved or unimproved actions could
be seen. It is only at larger coupling that deviations become
visible, and the improved lattice action in fact suppresses lattice
artifacts.

It is still an open problem to go to even stronger couplings.
Practical simulations become considerably more involved due to
stronger fluctuations in the sign of the fermion determinant.
Further refinements of our algorithm are already under
investigation, and we hope to report of our progress in the
near future. Apart from that, the attainable large statistics
allow for the determination of the (constrained) effective
potential for this theory; this might serve as an independent
check of the non-renormalization theorem for this particular
supersymmetric model.

We believe that a generalization of our numerical methods to
all supersymmetric theories without gauge fields can be
accomplished. In particular, the $\mathcal N=1$ model in
both two and four dimensions as well as supersymmetric
non-linear sigma models are within reach. At least the
experience gained in two-dimensional models suggests that
SLAC and twisted Wilson fermions might be good candidates for
the formulation of four-dimensional supersymmetric lattice
theories.

\begin{acknowledgments}
\noindent
We thank S.~D\"urr for conversations about the determination of masses. Further
we thank P.~Gerhold and K.~Jansen for helpful
discussions concerning algorithmic details and Fourier acceleration.
TK acknowledges support by the Konrad-Adenauer-Stiftung, GB by the Evangelisches Studienwerk and
CW by the Studienstiftung des deutschen Volkes. This work has been supported by
the DFG grant Wi~777/8-2.
\end{acknowledgments}


\appendix

\section{Determination of masses from two-point correlators}
\label{app:masses}
\noindent
One important observable of a quantum field theory is the energy gap between the
ground state and the first excited state. This energy gap corresponds to the
mass of the lightest particle in the spectrum.

To obtain the masses in the Wess-Zumino model one has to consider the propagators of fermions and bosons. 
At vanishing spatial momentum $p_1=0$, the free bosonic continuum propagator in momentum space
reads
\begin{equation}
G^\text{boson}(p) = \frac{1}{m^2+p_0^2}\, .
\end{equation}
The real and imaginary parts $\varphi_1$ and $\varphi_2$ of $\varphi$
decouple (the propagator is diagonal and even equal for $\varphi_1$, $\varphi_2$).
The Fourier transform of $G^\text{boson}(p)$ shows the well known exponential decay
\begin{equation}\label{eq:expFalloff}
C^\text{boson}(t) \propto \exp(-m\abs{t})\, ,
\end{equation}
where $m$ is the above mentioned mass of the lightest particle.
(The space coordinates corresponding to $p_1$ and $p_0$ are called $x$ and $t$, respectively.)
In the interacting case this quantity can be obtained on the lattice by measuring the two-point function.
The projection onto $p_1=0$ can be achieved by averaging over the spatial lattice sites,
\begin{equation}
C^\text{boson}_{\alpha\beta}(t) = \frac{1}{\Ns}\sum_{x}\vev{\varphi_\alpha(0,0)\varphi_\beta(t,x)}\, ,
\end{equation}
with $\alpha$, $\beta$ labeling components of the bosonic field.

The free fermionic continuum correlator for $p_1=0$ is
\begin{equation}
\vev{\psi_{\alpha}\bar{\psi}_{\beta}}=G^\text{fermion}_{\alpha\beta}(p_0) = \frac{m-i\gamma^0_{\alpha\beta}p_0}{m^2+p_0^2}.
\end{equation}
Using the representation of the $\gamma$ matrices as described after \eqref{eq:wz2d:nic7}
one can read off a direct connection with the bosonic correlator using
\begin{equation}
\label{eq:freeconnect}
G^\text{fermion}(p_0) := G^\text{fermion}_{11}(p_0)+G^\text{fermion}_{22}(p_0) = \frac{2m}{m^2+p_0^2}.
\end{equation}
As in the bosonic case on the lattice a summation over the spatial lattice sites yields the projection onto $p_1=0$.
$C^\text{fermion}(t)$ defines the Fourier transform of this object.
\subsection{Fermion masses}
\label{ssec:twopoint:fermions}
\noindent
The fermionic propagator $C(x)$ is given by 
\begin{equation}
 \vev{\psi_{\alpha}\bar{\psi}_{\beta}}=\vev{M^{-1}_{\alpha\beta}(\varphi_1,\varphi_2)}\, ,
\end{equation}
where $M$ is the fermion matrix. The calculation of this quantity
requires a high numerical effort for the inversion of large matrices. 
Fortunately in the weak-coupling limit the fermion
matrix is approximately the same as that of the free theory and the statistical
fluctuations are rather small. Therefore the necessary statistics to read off a reasonable
fermionic correlator is much smaller than for bosons.

After the fermionic correlator in position space is computed the masses can be
determined from its long range behavior. Inspired by the continuum connection between fermionic and bosonic correlators, \eqref{eq:freeconnect}, and the behavior at large distances,
\eqref{eq:expFalloff}, one can consider
\begin{equation}
\label{eq:logmeff}
\meff = \ln\left(\frac{C^\text{fermion}(t)}{C^\text{fermion}(t+1)}\right)
\end{equation}
with $t$ in a region between zero and $\Nt/2$. The mass can then be determined from the average of $m_{\rm eff}$. 

A more elaborate way is a least square fit of the fermionic correlator
$C^\text{fermion}(t)$ with the function
\begin{equation}
\label{eq:coshmass}
f_{a,m_\text{f}}(t) = a\cdot \cosh (m_\text{f}(t-\Nt/2))
\end{equation}
One better not take the full range of $t$ into account for this fit because it is valid only for large distances (for periodic boundary conditions, from both boundaries of the lattice). One should therefore constrain $t$ to be in 
$\{1+\tskip,\ldots,\Nt-1-\tskip\}$.
The choice of $\tskip$ is determined by the fringe of the plateau in a plot of the fitting result vs.\ $\tskip$.

\begin{figure}
\input{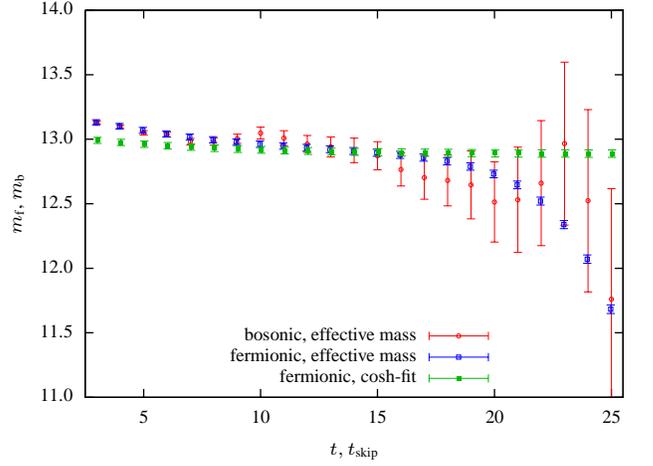}
\caption{\label{fig:massPlateauWilson}
Bosonic and fermionic masses obtained via a $\cosh$-fit \eqref{eq:coshmass} and the effective mass
definition \eqref{eq:logmeff}
for the improved Wilson model with $\lambda=0.4$ on a $64\times 64$ lattice. The
fermionic masses with
a statistics of about 5,000 independent configurations are much sharper and
more reliable than the bosonic effective masses obtained from about $10^6$ independent
configurations.}
\end{figure}

The differences of the different methods to determine the masses are illustrated in Fig.~\ref{fig:massPlateauWilson}.
One clearly observes that the effective masses determined according to \eqref{eq:logmeff} do not show a plateau from
which the mass can be read off. By contrast, the masses obtained from a
$\cosh$ fit clearly show this behavior at large $\tskip$. As mentioned above,
the effective mass of the bosonic correlator is subject to much larger statistical errors.

\subsection{Boson masses}
\noindent
In order to calculate the bosonic correlators for the determination of the masses the connected
two-point function is considered.
At large distances, where the masses can be extracted, the relative statistical error of the correlator grows exponentially.
Therefore, one must achieve a balance between this statistical error and the systematical errors due to the evaluation at small distances.

We have fitted $\ln(C^\text{boson}(t))$ against the function $A+\ln(\cosh(m_\text{b}/\Ns(t-\Nt/2)))$ to determine $A$ and the effective mass $m_\text{b}$.
In order to exclude the points with the largest statistical and systematical errors from this fit, we have taken only the points in the interval ($[\tskip,t_\text{st}]\cup[N_t-t_\text{st},N_t-\tskip]$) into account. $\tskip$ is determined as in the fermionic case and $t_\text{st}$ such that the  statistical error becomes comparably small.

If the SLAC derivative is used an oscillatory behavior of $m_\text{b}$ as a function of $\tskip$ can be observed. In the bosonic case it is slightly smaller than the statistical error.
Therefore, it is sufficient to measure a ``smeared'' mass,
$m_\text{SLAC} = 0.5 m_\text{b}(\tskip,t_\text{st})+0.25 m_\text{b}(\tskip+1,t_\text{st}) + 0.25 m_\text{b}(\tskip-1,t_\text{st})$, where the error of the oscillations is negligible as compared to the statistical one.

\section{Continuum extrapolation}
\label{app:contExtr}
\noindent
For the continuum extrapolation we focus on  the fermionic masses because of their much smaller statistical error. 
The explicit extrapolation procedure is guided by analytic results and observations for the free theory.
The three different discretizations investigated in this work require different strategies for this procedure. 
\subsection{Wilson derivative}
\noindent
Compared with the continuum formula, \eqref{eq:freeconnect}, the free momentum space correlation function
for the Wilson derivative gets a momentum dependent mass,
\begin{equation}
G^\text{fermion}(p_0) = \frac{\mlatt+1-\cos(p_0)}{\sin^2(p_0)+(\mlatt+1-\cos(p_0))^2} .
\end{equation}
The pole of this correlator coincides with the above mentioned $\cosh$-fit within the error bars.

To extrapolate the continuum limit an expansion in powers of the lattice spacing is used. 
Exact results for the free theory were derived to check this extrapolation. 
In this case an expansion up to a linear order in $a$ is not enough to obtain
the known result within the high precision of the numerical measurements at weak
coupling. Therefore we first tried to extended the expansion to a quadratic
order
 which yields a better result; but still the error is to large for our purposes.

The functional behavior of the masses, $m_\text{f}$, obtained by the fit as a function of the lattice spacing is well approximated by 
\begin{equation}\label{eq:wilsonMassExtrap}
m_f(a) \approx m_\cont + A\cdot a + B\cdot a^{\frac{3}{2}}
\end{equation}
for all $a\in[0,0.05]$.
The deviation from this behavior is negligible with respect to the statistical
errors in the weak coupling case.
In addition the expected continuum result is achieved with the necessary precision.
Motivated by these results this formula is also used in the interacting case.
\subsection{Twisted Wilson derivative}
\noindent
A Wilson parameter of $r=\sqrt{\frac{4}{3}}$ for the twisted Wilson fermions
in the free theory leads to discretization errors of $\ord(a^4)$ as discussed
in \cite{Bergner:2007pu}. For the weakly coupled regime ($\lambda\le 0.3$) we
expect these errors to dominate the lattice artifacts. Nevertheless for an
intermediate coupling corrections of $\ord(a)$ arise. Taking this into account we
extrapolate the masses to continuum assuming a functional behavior of
\begin{equation}
m_f(a) = m_\cont + A\cdot a + B\cdot a^4.
\end{equation}
For $\lambda>0.3$ the $\ord(a)$ terms dominate. Therefore a
linear extrapolation is sufficient.
\subsection{SLAC derivative}
\begin{figure}[b]
   \input{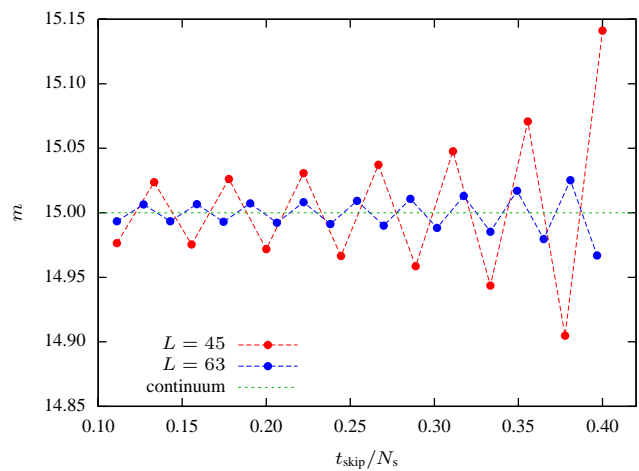}
\caption{\label{fig:slacFreeMass}
Masses obtained via a $\cosh$-fit for the free theory using the SLAC derivative.
At larger lattices the oscillation amplitude around the continuum value gets smaller.}
\end{figure}

\noindent
As we have seen in our previous investigations,  \cite{Bergner:2007pu}, the SLAC-derivative shows an almost perfect behavior. That means the extrapolated masses coincide with their continuum counterparts already at finite lattice spacings.
On the other hand we have observed an oscillatory behavior of the correlation function. 
This was shown to be connected with the exact reproduction of the continuum dispersion relation by the SLAC derivative. 
To handle this problem we have again studied the free theory first. As in the
bosonic case  the plot of $m_\text{f}$ versus $\tskip$ does  not show a clear plateau
but rather oscillates around the expected continuum value, cf.~Fig.~\ref{fig:slacFreeMass}.

Guided by these observation of the free theory a suitable averaging can lead to
the extraction of the correct continuum results at finite lattice spacing. Starting with the ansatz
\begin{equation}
m(\Ns,\bm{c}) := c_0 m_\text{f}(\tskip)+c_1 m_\text{f}(\tskip-1)+c_2 m_\text{f}(\tskip-2).
\end{equation}
we minimize the difference form the known continuum result of the free theory
\begin{equation}
\Delta(\Ns,\bm{c}) = \abs{m(\Ns,\bm{c})-m_\cont}
\end{equation}
for lattice sizes of $\Ns=\Nt\in\{35,37,\ldots,75\}$ and $\tskip=\left\lfloor 
0.4\Ns\right\rfloor$. A least square fit yields
\begin{equation}
c_0 = 0.11791,\quad c_1 = 0.47877,\quad c_2 = 0.40332\, ,
\end{equation}
leading to $\max \Delta(\Ns,\bm{c}) =
5.282\times 10^{-4}$.  A smaller $\tskip$ does not change this result considerably.
Using this approximation scheme the systematic error based on the oscillatory
behavior of the SLAC derivative can be neglected compared to the
statistical errors at least for the weak coupling case.
\renewcommand{\eprint}[1]{ \url{[arXiv:#1]}}

\end{document}